

\magnification=1200 \hfuzz=10pt \font\male=cmr8
\font\titlefont=cmbx10 scaled\magstep1


\def\r{representation ~} 
\def\l{\Lambda_\chi} \def\cg{{\cal G}} \def\ch{{\cal H}}
\def\cc{{\cal C}}\def\cu{{\cal U}} 
  
\def\bc{{\bf C}} \def\vr{\vert} \def\L{\Lambda}
 \def\Us{\cu_q(su(2,2))} \def\om{\omega}
\def\a{\alpha} \def\d{\delta}  
 \def\g{\gamma} \def\ve{\varepsilon} \def\bn{{\bf
N}}   


\line{\hfill IC/92/187} \line{\hfill SISSA 137/92/FM}
\line{\hfill UTS-DFT-22-92} \line{\hfill July 1992} \vskip 1.5cm

\centerline{\titlefont POSITIVE ENERGY REPRESENTATIONS}
\smallskip \centerline{\titlefont OF THE CONFORMAL QUANTUM
ALGEBRA} \vskip 1.5cm

\centerline{{\bf L. D\c abrowski$^\star$}, {\bf V.K.
Dobrev$^{+*}$}, {\bf R. Floreanini$^\dagger$} and {\bf V.
Husain$^{+**}$}} \footnote{}{$^{*}$ ~\male{Permanent address and
after 31 July 1992 : Bulgarian Academy of Sciences, Institute of
Nuclear Research and Nuclear Energy, 72 Tsarigradsko Chaussee,
1784 Sofia, Bulgaria.}\hfil\break\indent  $^{**}$ ~\male{Address
after 30 September 1992 : Theoretical Physics Institute,
University of Alberta, Edmonton, Canada T6G 2J1.}}

\vskip 1cm

\item{$^\star$} International School for Advanced Studies,
SISSA, Trieste, Italy \item{$^+$} International Center for
Theoretical Physics, Trieste, Italy \item{$^\dagger$} Istituto
Nazionale di Fisica Nucleare, Sezione di Trieste, Dipartimento di
Fisica \hbox{Teo-}\hfil\break rica, Universit\`a di Trieste,
Strada Costiera 11, 34014 Trieste, Italy

\vskip 2cm

\centerline{\bf Abstract}
\smallskip\midinsert\narrower\narrower\noindent The
positive-energy unitary irreducible representations of the
$q$-deformed conformal algebra $\cc_q = $ ${\cal U}_q(su(2,2))$
are obtained by appropriate deformation of the classical ones.
When the deformation parameter $q$ is $N$-th root of unity, all
these unitary representations become finite-dimensional. For this
case we discuss in some detail the massless representations,
which are also irreducible representations of the $q$-deformed
Poincar\'e subalgebra of $\cc_q$. Generically, their dimensions
are smaller than the corresponding finite-dimensional non-unitary
representation of $su(2,2)$, except when $N=2$, $h=0$ and $N = 2
\vr h\vr +1$, where $h$ is the helicity of the representations.
The latter cases include the fundamental representations with $h
= \pm 1/2$.

\endinsert

\vfill\eject

\null

{\bf 1.} ~The positive energy unitary irreducible representations
(UIRs) of the conformal algebra $su(2,2)$ of four dimensional
Minkowski space are certainly of physical relevance. They are
given in Ref.[1]. In this letter we consider the $q$-deformed
conformal algebra ${\cal C}_q\equiv {\cal U}_q(su(2,2))$ [2] and
present the $q$-analogs of the UIRs of [1].

For generic complex $q$, the positive energy UIRs $V_q$ of
$\cc_q$ are deformations of the respective representations of
$su(2,2)$. Here, our considerations amount to an introduction of
the correct scalar product in the representations $V_q$. It turns
out that we can use the same scalar product as for the undeformed
representations. However, we have to extend the hermitian
conjugation used in [1] to a hermitian conjugation of the
complexification ${\cal U}_q(sl(4,\bc))$ of ${\cal
U}_q(su(2,2))$. For $|q| = 1$, this is done in Section 3.

When the deformation parameter $q$ is a root of unity, the
picture of the representations changes drastically. All
irreducible representations of ${\cal U}_q(su(2,2))$, as
inherited from ${\cal U}_q(sl(4,\bc))$, are finite-dimensional.
In Section 4 we give a list of the positive energy UIRs. We
emphasize that, unlike the classical case, these unitary
representations are finite-dimensional. We discuss in more detail
the massless finite-dimensional representations of $\cc_q$; these
are also UIRs of the $q$-deformed Poincar\'e subalgebra of
$\cc_q$.

\bigskip

{\bf 2.} ~The physically relevant representations of the
4-dimensional conformal algebra $su(2,2)$ may be labelled by
$\chi = [j_1,j_2,d]$, where $2j_1, 2j_2$ are non-negative
integers fixing finite dimensional irreducible representations of
the Lorentz subalgebra $so(3,1)$, and ~$d$~ is the conformal
dimension (or energy). We would like to label the representations
of ${\cal U}_q(su(2,2))$ by the same set of indices $\chi$ and
because of this we shall use the $q$-deformed conformal algebra
${\cal U}_q(su(2,2))$ (cf. [2]) which has the $q$-deformed
Lorentz algebra $\cu_q(so(3,1))$ as a Hopf subalgebra. Since
${\cal U}_q(su(2,2))$ is a real from of ${\cal U}_q(sl(4,{\bf
C}))$ we need first to recall the latter deformation.

The $q$-deformation ${\cal U}_q(sl(4,{\bf C}))$ of the universal
enveloping algebra ${\cal U}(sl(4,{\bf C}))$ is defined [3,4] as
the associative algebra over $\bc$ with Chevalley generators
$X^\pm_j ~, ~H_j ~, ~j = 1,2,3$ and with relations~: $$[H_j~,
{}~H_k] ~ = ~ 0 , ~~[H_j~, ~X^\pm_k] ~ = ~ \pm a_{jk} X^\pm_k ~,
\eqno(1a)$$ $$[X^+_j ~, ~X^-_k] ~ = ~ \d_{jk} {q^{H_j/2} -
q^{-H_j/2} \over q^{1/2} - q^{-1/2}} ~ = ~ \d_{jk} ~[H_j]_q ~,
\eqno(1b)$$ $$\eqalign{ \left( X^\pm_j \right)^2 X^\pm_k - [2]_q
X^\pm_j X^\pm_k X^\pm_j + X^\pm_k \left( X^\pm_j \right)^2 ~& = ~
0 , ~~(jk) = (12),(21),(23),(32) \cr [ X^\pm_1 , X^\pm_3 ] ~& = ~
0 ~, \cr }\eqno(1c)$$ where $ [x]_q = \bigl( q^{x/2} - q^{-x/2}
\bigr) / \bigl( q^{1/2} - q^{-1/2} \bigr) $, $(a_{jk}) = (2(\a_j
, \a_k)/(\a_j , \a_j))$, $j,k = 1,2,3$, is the Cartan matrix of
$sl(4,{\bf C})$; $\a_1,\a_2,\a_3$ are the simple roots; the
non-zero products between the simple roots are: $(\a_j,\a_j) =
2$, $j = 1,2,3$, $(\a_1,\a_2) = (\a_2,\a_3) = -1$. The non-simple
positive roots are : $\a_{12} = \a_1+\a_2$ ~ , $\a_{23} =
\a_2+\a_3$ ~ , $\a_{13} = \a_1 + \a_2 + \a_3$. The elements $H_j$
span the Cartan subalgebra $\ch$ while the elements $X^\pm_j$
generate the subalgebras $\cg^\pm$.

The algebra $\cu_q(sl(4,\bc))$ is a Hopf algebra [5] with
co-multiplication $\d$, co-unit $\ve$ (homomorphisms) and
antipode $\g$ (antihomomorphism) defined on the generators as
follows [3,4]: $$\d(H_j) ~ = ~ H_j \otimes 1 ~+~ 1\otimes H_j ~,
{}~~~\d(X^\pm_j) ~ = ~ X^\pm_j\otimes q^{H_j/4} ~+~
q^{-H_j/4}\otimes X^\pm_j ~, \eqno(2a)$$ $$\ve(H_j) ~ = ~
\ve(X^\pm_j) ~ = ~ 0 ~, ~~~~\g(H_j) ~ = ~ -H_j ~, ~~~\g(X^\pm_j)
{}~ = ~ -q^{\pm 1/2}~X^\pm_j ~. \eqno(2b)$$

The Cartan-Weyl basis for the non-simple roots is given by (cf.
[4,6]): $$X^\pm_{jk} ~ = ~ \pm (q^{1/4} X^\pm_j X^\pm_k -
q^{-1/4} X^\pm_k X^\pm_j ) ~, ~~(jk) ~ = ~ (12), (23) ~,
\eqno(3a)$$ $$\eqalign{ X^\pm_{13} ~& = ~ \pm
(q^{1/4} X^\pm_1 X^\pm_{23} - q^{-1/4} X^\pm_{23} X^\pm_1 ) ~ =
\cr & = ~ \pm (q^{1/4} X^\pm_{12} X^\pm_3 - q^{-1/4} X^\pm_3
X^\pm_{12} ) ~. \cr } \eqno(3b)$$ All other commutation relations
and Hopf algebra relations for the generators in $(3)$ follow
from these definitions.

Let us consider the conformal algebra $su(2,2) \cong so(4,2)$. It
has 15 generators $Y_{AB} = - Y_{BA}$, $A,B = 1,2,3,5,6,0$
satisfying $$ [Y_{AB},Y_{CD}] ~ = ~ \eta_{BC} Y_{AD} - \eta_{AC}
Y_{BD} - \eta_{BD} Y_{AC} +\eta_{AD} Y_{BC} \ , \eqno(4)$$ where
$\eta_{AB} = $ diag$(----++)$. Since $su(2,2)$ is the conformal
algebra of 4-dimensional Minkowski space-time we use a
deformation [2] consistent with the subalgebra structure relevant
for the physical applications. These subalgebras are: the {\it
Lorentz} subalgebra $so(3,1)$ generated by $Y_{\mu
\nu}$, $\mu, \nu = 1,2,3,0$, the subalgebra of {\it translations}
generated by $P_\mu = Y_{\mu 5} + Y_{\mu 6}$, the subalgebra of
{\it special conformal transformations} generated by $K_\mu =
Y_{\mu 5} - Y_{\mu 6}$, the {\it dilatations} subalgebra
generated by $D = Y_{56}$. For a deformation of $su(2,2)$ one
uses the expressions for its generators as complex linear
combinations of the generators of its complexification
$sl(4,\bc)$ compatible with  the reality structure. The
deformation ${\cal U}_q(su(2,2))$ introduced in [2] uses
essentially the same linear combinations as in [1] (cf. (2.21))
and we omit these explicit expressions for the lack of space.

\bigskip

{\bf 3.} ~In this section we consider the representations of
$\cc_q = {\cal U}_q(su(2,2))$ in the generic case when the
deformation parameter is not a root of unity. In this case the
representations of $\cc_q$ we use are irreducible lowest weight
modules $M^\chi$ (in particular, Verma modules $V^\chi$) of
${\cal U}_q(sl(4,\bc))$ together with the reality condition
necessary for the construction of the scalar product in $M^\chi$
(or $V^\chi$).

We use the standard decomposition $\cg = sl(4,\bc) = \cg^+ \oplus
\ch \oplus \cg^-$, where $\ch$ and $\cg^\pm$ were introduced in
Section 2. A lowest weight module $M^\chi$ of ${\cal
U}_q(sl(4,\bc))$ is given by its lowest weight $\L_\chi \in
\ch^*$ (${\cal H}^*$ being the dual of $\cal H$) and lowest
weight vector $v_0\equiv v_0(\chi)$, such that $v_0$ is
annihilated by the lowering generators, $Xv_0 = 0$, $X \in {\cal
U}_q(\cg^-)$, and $ Hv_0 = \Lambda_\chi(H) v_0$ for any Cartan
generator $H$.

In particular, we use the Verma modules $V^\chi$ which are the
lowest weight modules such that $V^\chi = {\cal U}_q(\cg^+)\,
v_0$. Thus the Poincar\'e-Birkhof-Witt theorem tells us that the
basis of $V^\chi$ consists of monomial vectors $$ \Psi_{\{k\}} =
(X_1^+)^{k_1} (X_2^+)^{k_2} (X_3^+)^{k_3} (X_{12}^+)^{k_{12}}
(X_{23}^+)^{k_{23}} (X_{13}^+)^{k_{13}}\, v_0\, \quad k_j,
k_{jk}\in {\bf Z}_+\ ,\eqno(5)$$ where $X^+_j\equiv X^+_{jj}$ and
$X_{jk}^+$ are the six raising generators of $sl(4,{\bf C})$ (cf.
Section 2). In order to consider $V^\chi$ as a representation of
the real form it is not enough to express the generators of
$\cu_q(sl(4,\bc))$ as linear combinations of generators of ${\cal
U}_q(su(2,2))$. We have to introduce, as in the $q = 1$ case, a
hermiticity condition invariant with respect to ${\cal
U}_q(su(2,2))$. Such a condition is well known in the undeformed
case and is given by (cf. [1] (4.8)):
$$\eqalignno{&\omega(X_{jk}^{\pm}) = \cases{X^\mp_{jk}\ ,& ($jk$)
= (11), (33)\cr -X^\mp_{jk}\ ,& otherwise\ ,\cr} &(6a)\cr
&\omega(H) = H\ , \qquad \forall H\in \ch\ . &(6b)\cr}$$  The
problem in the $q$-deformed case is to extend this conjugation to
${\cal U}_q(sl(4,\bc))$ as an anti-linear anti-involution thus
making ${\cal U}_q(sl(4,\bc))$ a $*$-Hopf algebra. For this
extension it is enough to postulate (6) for the simple root
vectors and the corresponding Cartan subalgebra elements, i.e.,
for $X_k^\pm\equiv X_{kk}^\pm$ and $H_k$ for $k = 1,2,3$. Then
(6) follows for the non-simple root vectors. Indeed, take for
example $X_{12}^+\equiv q^{1/4} X_1^+X_2^+ - q^{-1/4} X_2^ +
X_1^+$. Then we have (iff $\vr q \vr = 1$): $$\eqalign{\om
(X_{12}^+)\ = &\ {\bar q^{1/4}} \om ( X_2^+) \om (X_1^+) - {\bar
q^{-1/4}} \om (X_1^+)\om (X_2^+)\ = \ - {\bar q^{1/4}} X_2^-
X_1^- + {\bar q^{-1/4}} X_1^- X_2^- \ = \cr = &\ - (q^{-1/4}
X_2^-X_1^- - q^{1/4} X_1^- X_2^-) \ = \ - X_{12}^- \ . \cr} $$
The same considerations go for the other non-simple root vectors.
For the other commutation relations $\om $ acts as an anti-linear
anti-involution. Finally one can check that $\om $ is an
anti-linear coalgebra anti-involution. Thus ${\cal
U}_q(sl(4,\bc))$ is a $*$ - Hopf algebra with $\vr q \vr = 1$.

As in the undeformed case, the conjugation $\om $ will be used to
introduce a $\Us$-invariant scalar product, i.e., a scalar
product such that the generators $X$ of $\cc_q$ are skew
hermitian. As in the undeformed case (cf. [1] (2.17)) it is
convenient to make a basis transformation so that the generators
of $\cc_q$ obey: $\om(X) = -X$. A set of  such generators is
given by:  $$\eqalign{&iH_k, ~k=1,2,3, ~~~~X^+_k - X^-_k,
{}~~i(X^+_k + X^-_k), ~k=1,3, \cr &X^+_{jk} + X^-_{jk},
{}~~i(X^+_{jk} - X^-_{jk}), ~~(jk) = (22), (12), (23), (13) ~.
\cr}\eqno(7)$$ In particular, the conformal Hamiltonian ~$H_0$~
[1] is given in the two bases (4) and (7) by:  $$H_0 ~=~ {1 \over
2} (P_0 + K_0) ~=~ Y_{05} ~=~ {1 \over 2} (H_1 + H_3) + H_2 ~.
\eqno(8)$$

For the scalar product of two vectors of the form (5) we take, as
in [1], $$\eqalign{ (\Psi_{\{k^\prime\}},
\Psi_{\{k\}}) & = \Bigl( v_0^*,\, \om \bigl( (X_{13}^+)^{k'_{13}}
\bigr) \om \bigl( (X_{23}^+)^{k'_{23}} \bigr) \om \bigl(
(X_{12}^+)^{k'_{12}} \bigr) \om \bigl( (X_3^+)^{k'_3} \bigr) \om
\bigl( (X_2^+)^{k'_2} \bigr) \cr &\times\om \bigl( (X_1^+)^{k'_1}
\bigr) (X_1^+)^{k_1} (X_2^+)^{k_2} (X_3^+)^{k_3}
(X_{12}^+)^{k_{12}} (X_{23}^+)^{k_{23}} (X_{13}^+)^{k_{13}}\, v_0
\Bigr) \cr & = (-1)^{k'_{13} + k'_{23} + k'_{12} + k'_{2}} \Bigl(
v_0^*,\, (X_{13}^-)^{k'_{13}} (X_{23}^-)^{k'_{23}}
(X_{12}^-)^{k'_{12}}(X_3^-)^{k'_3} (X_2^-)^{k'_2} \cr &\times
(X_1^-)^{k'_1}(X_1^+)^{k_1} (X_2^+)^{k_2} (X_3^+)^{k_3}
(X_{12}^+)^{k_{12}} (X_{23}^+)^{k_{23}} (X_{13}^+)^{k_{13}}\, v_0
\Bigr) \ ,\cr} \eqno(9)$$ with $(v_0^*,v_0) = 1 $. (Note that (9)
is an adaptation of the classical contravariant Schapovalov
form.) Calculation of (9) is performed in the standard manner by
moving the lowering (raising) generators to the right (left)
where they annihilate $v_0\ (v_0^*)$. Finally, the result is some
polynomial in $\L_\chi(H_{j})$. Thus we have to specify how
$\L_\chi(H)$ is fixed by the \r $\chi$ which we take as in [1],
{\it i.e.} we have $$ \l(H_1) = -2j_1\ , \qquad \l(H_2) =
d+j_1+j_2\ ,\qquad \l(H_3) = -2j_2\ . \eqno(10) $$ Note that our
generators $H_1,\ H_2,\ H_3$ correspond to $2H_1,\ H_0-H_1-H_2, \
2H_2$, respectively, of [1]. In particular, using (8) we see that
{}~$d$~ is the eigenvalue of the conformal Hamiltonian $H_0$; hence
$d$ is called the conformal energy (or dimension).

Given the scalar product (9), we have to determine whether it
provides an UIR of ${\cal U}_q(su(2,2))$. It is clear that the
conditions on $j_1$, $j_2$ and $d$ will be the same as in [1],
and below we specify precisely the UIR spaces applying new
results in the representation theory of Verma modules ([6] and
references therein).

Generically, the Verma modules $V^\chi$ are irreducible. A Verma
module $V^\chi$ is reducible [6] iff there exists a positive root
$\alpha$ and a positive integer $m_\a$ such that the following
equality holds $$ \bigl[(\L_\chi-\rho)(H_\alpha) + m_\alpha
\bigr]_q = \, 0\ ,\eqno(11)$$ where $H_\alpha$ is a linear
combination of $H_k$, specifically, if $\alpha = \sum_k n_k
\alpha_k$, $n_k\in {\bf Z}_+$, $\alpha_k$ are simple roots, then
$H_\alpha = \sum_k n_k H_k$, and $\rho$ is half the sum of
positive roots; note that $\rho(H_k) = 1$. For the six positive
roots of the root system of $sl(4,{\bf C})$, one has (see
Ref.[7]): $$\eqalignno{ &m_1 = - \L_\chi(H_1) + 1 = 2 j_1 + 1\ ,
&(12a)\cr &m_2 = - \L_\chi(H_2) + 1 = 1 - d - j_1 - j_2\ ,
&(12b)\cr &m_3 = - \L_\chi(H_3) + 1 = 2 j_2 + 1\ , &(12c)\cr
&m_{12} = - \L_\chi(H_{12}) + 2 = m_1 + m_2 = 2 - d + j_1 - j_2\
, &(12d)\cr &m_{23} = - \L_\chi(H_{23}) + 2 = m_2 + m_3 = 2 - d -
j_1 + j_2\ , &(12e)\cr &m_{13} = - \L_\chi(H_{13}) + 3 = m_1 +
m_2 + m_3 = 3 - d + j_1 + j_2\ . &(12f)\cr}$$ Whenever (11) is
fulfilled there exists a singular (null) vector $v_s$ in $V^\chi$
such that $v_s\neq v_0$, $X v_s = \, 0$, $X\in {\cal U}_q(\cg^-)$
and $H_\alpha v_s = (\L_\chi + m_\alpha)(H_\a)\, v_s$.

To obtain an irreducible lowest weight module we have to factor
out all singular vectors. First of all, we have that $m_1$ and
$m_3$ are positive, since $2j_1$ and $2j_2$ are non-negative
integers. The corresponding singular vectors are $$v_1\ = \
\bigl( X_1^+ \bigr)^{2j_1+1} v_0\ , \ \ v_3\ = \ \bigl( X_3^+
\bigr)^{2j_2+1} v_0\ , \eqno(13)$$ and these are present for all
representations we discuss. Next, it is clear that depending on
the value of $d$ there may be other singular vectors. Since we
are interested in the positive-energy UIRs, we recall the list of
these representations for $su(2,2)$ [1]: $$\eqalign{ & 1)\ \
d>j_1+j_2+2\ ,\qquad j_1 j_2\neq 0\ ,\cr & 2)\ \ d = j_1+j_2+2\
,\qquad j_1 j_2\neq 0\ ,\cr & 3)\ \ d>j_1+j_2+1\ ,\qquad j_1 j_2
= \, 0\ ,\cr & 4)\ \ d = j_1+j_2+1\ ,\qquad j_1 j_2 = \, 0\ ,\cr
& 5)\ \ d = j_1 = j_2 = \, 0\ .\cr} \eqno(14)$$ As we have
already said, the same list is valid for ${\cal U}_q(su(2,2))$.
In case 1) there are no additional singular vectors. If $d =
j_1+j_2+2$, which is case 2) and is also possible in case 3),
then $m_{13} = 1$ and there is an additional singular vector:
$$\eqalign{v^{(1)}_{13} ~& = ~ \Bigl( [2j_1] [2j_2] X^+_1 X^+_3
X^+_2 ~-~ [2j_1] [2j_2 +1] X^+_1 X^+_2 X^+_3 ~- \cr &-~ [2j_1+1]
[2j_2] X^+_3 X^+_2 X^+_1 ~+~ [2j_1+1] [2j_2+1] X^+_2 X^+_1 X^+_3
\Bigr) ~v_0 ~, \cr }\eqno(15)$$ where $[x] = [x]_q = \bigl(
q^{x/2}-q^{-x/2} \bigr) / \bigl( q^{1/2}-q^{-1/2} \bigr) $. In
case 4) we have $m_{13} = 2$. Moreover, $m_{23} = 1$ if $j_1 = 0$
and $m_{12} = 1$ if $j_2 = 0$. Thus there are two singular
vectors if $j_1+j_2>0$, and three singular vectors if $j_1 = j_2
= \, 0$. These vectors are $$\eqalign{ v^{(2)}_{13} ~& = ~
\Bigl( [2j_1] [2j_1-1] [2j_2] [2j_2 -1] (X^+_1)^2 (X^+_3)^2
(X^+_2)^2 ~-\cr &~ - [2] [2j_1] [2j_1-1] [2j_2+1] [2j_2 -1]
(X^+_1)^2 X^+_3 (X^+_2)^2 X^+_3 ~+ \cr &~ + [2j_1] [2j_1-1]
[2j_2+1] [2j_2] (X^+_1)^2 (X^+_2)^2 (X^+_3)^2 ~- \cr &~ - [2]
[2j_1+1] [2j_1-1] [2j_2] [2j_2 -1] X^+_1 (X^+_3)^2 (X^+_2)^2
X^+_1 ~+ \cr &~ + [2]^2 [2j_1+1] [2j_1-1] [2j_2+1] [2j_2 -1]
X^+_1 X^+_3 (X^+_2)^2 X^+_3 X^+_1~- \cr &~ - [2] [2j_1+1]
[2j_1-1] [2j_2+1] [2j_2] X^+_1 (X^+_2)^2 (X^+_3)^2 X^+_1 ~+ \cr
&~ + [2j_1+1] [2j_1] [2j_2] [2j_2 -1] (X^+_3)^2 (X^+_2)^2
(X^+_1)^2 ~- \cr &~ - [2] [2j_1+1] [2j_1] [2j_2+1] [2j_2 -1]
X^+_3 (X^+_2)^2 X^+_3 (X^+_1)^2 ~+ \cr &~ + [2j_1+1] [2j_1]
[2j_2+1] [2j_2] (X^+_2)^2 (X^+_3)^2 (X^+_1)^2 \Bigr) ~v_0 ~, \cr
&~ d = j_1 + j_2 + 1 ~, ~~j_1 j_2 = 0 ~, ~m_{13} = 2 ~, \cr }
\eqno(16a)$$ $$\eqalignno{v_{12} ~& = ~ \bigl( [2j_1] X^+_1
X^+_2 - [2j_1 + 1]X^+_2 X^+_1 \bigr) ~ v_0 ~, ~~~ d = j_1 + 1 ~,
{}~~ j_2 = 0 ~, ~m_{12} = 1 ~, &(16b)\cr v_{23} ~& = ~ \bigl(
[2j_2] X^+_3 X^+_2 - [2j_2 + 1] X^+_2 X^+_3 \bigr) ~ v_0 ~, ~~~ d
= j_2 + 1 ~, ~~ j_1 = 0 ~, ~m_{23} = 1 ~, &(16c)\cr }$$ and all
expressions in (16) are valid also when $j_1 = j_2 = \, 0$.
Finally, in case 5), $m_1 = m_2 = m_3 = 1$, there are three
singular vectors $X_k^+\, v_0$, $k = 1,2,3$, and when factored
out, the whole ${\cal U}_q(\cg^+)$ gives zero contribution,
yielding the one-dimensional representation. Factoring out all
the singular vectors together with their descendents, one can now
explicitly build the positive-energy representations.

\bigskip

{\bf 4.} ~In this section we consider the case where the
deformation parameter is a root of unity, namely, $q = e^{2\pi
i/N}$, $N = 2,3,\ldots$

In this case all Verma modules $V^\chi$ are reducible [6] and all
irreducible representations are {\it finite dimensional} [8].
There are singular vectors for all positive roots $\alpha$ [6].
Condition (11) also has more content now because if
$(\Lambda_\chi - \rho)(H_\alpha) = -m\in {\bf Z}$, then (11) will
be fulfilled for all $m+kN$, $k\in {\bf Z}$. In particular, there
will be an infinite series of positive integers $m$ such that
(11) is true [6]. For identical reasons, there is an infinite
number of lowest weights $\l$ such that (11) is satisfied for the
same set of positive integers $m = m_\alpha$.

Let us take a representation such that for each simple root
$\alpha$ (11) is satisfied with a positive integer $m =
m_\alpha\le N$. Then the three numbers $m_1,m_2,m_3$ which
characterize the representation $V^\chi$ will be $$ \eqalign{&m_1
= \{2j_1 +1 \}_N\cr &m_2 = \cases{\{-d-j_1-j_2+1\}_N, &if
$d\in{\bf Z}$,\cr N, & if $d\not\in {\bf Z}$,\cr}\cr &m_3 =
\{2j_2+1\}_N\ ,\cr}\eqno(17)$$ where $\{x\}_N$ is the smallest
positive integer equal to $x$ (mod $N$); thus we have $m_k\in
\bn$ and $0< m_k \leq N$, $k = 1,2,3$.

The weights $\chi$ such that (17) is satisfied are divided into
six classes [9] depending on the values of $m_{12} = m_1+m_2$,
$m_{23} = m_2+m_3$, and $m_{13} = m_1+m_2+m_3$ : $$ \eqalignno{ &
a)\ \ m_{jk} \leq N\ , &(18a)\cr & b)\ \ m_{12},\, m_{23} \leq N\
,\quad N<m_{13}\leq 2N\ , &(18b)\cr & c)\ \ m_{12}\leq N\ ,\quad
N<m_{23},\, m_{13}\leq 2N\ , &(18c)\cr & d)\ \ m_{23}\leq N\
,\quad N<m_{12},\, m_{13}\leq 2N\ , &(18d)\cr & e)\ \ N<m_{12},\,
m_{13},\, m_{23}\leq 2N\ , &(18e)\cr & f)\ \ N<m_{12},\,
m_{23}\leq 2N\ ,\quad 2N<m_{13}\leq 3N\ . &(18f)\cr}$$ These
representations inherit all the structure from their ${\cal
U}_q(sl(4,\bc))$ counterparts. Thus the classification of the
unitarizable lowest weight representations of $\Us$ proceeds as
follows.

Imposing the conditions of positive energy UIRs (14), we see that
cases 1), 2), 3) are in one to one correspondence with the finite
dimensional representations of ${\cal U}_q(sl(4,\bc))$ for $q^N =
1$ and all possibilities listed in (18) are admissible. The same
list of representations is valid for ${\cal U}_q(su(4))$.

The classification of the massless case 4) of (14) is more
interesting since not all cases in (18) are admissible. Since
$j_1j_2 = 0$, let us choose for definiteness $j_2 = 0$. Then from
(17) we have $$ m_1 = \{2j_1+1\}_N\ ,\ \ \ m_2 = \{-2j_1\}_N\ ,\
\ \ m_3 = 1\ ,$$ $$ m_{12} = N+1\ ,\ \ \ m_{13} = N+2\ ,
\eqno(19)$$ $$m_{23} = \cases{\{1-2j_1\}_N\leq N\ , & if $j_1\neq
0$,\cr N+1\ , & if $j_1 = \, 0$.\cr}$$ Therefore the admissible
cases are $(18d)$ when $j_1\neq 0$, and $(18e)$ when $j_1 = 0$.
For the dimension $d(N,J_1)$ of these representations we have,
adapting a result of Ref.[9], $$ d(N,J_1) = {1\over 3} \Bigl[
2N^3-N(12J_1^2-1) + 3J_1(4J_1^2-1)\Bigr]\ , \eqno(20) $$ where
$J_1$ is such that $2J_1+1 = (2j_1+1)$ (mod $2N$) and $1\leq
2J_1+1\leq N$. We recall that in the classical case the massless
unitary representations are infinite-dimensional. However, we may
compare our representations with the undeformed non--unitary
finite--dimensional representations which have the same quantum
numbers $(m_1,m_2,m_3) = $ $(2J_1 +1, N-2J_1 ,1)$. We note that
the dimension of the former is generically smaller than the
dimension of the latter, which is given by $$d_c(m_1,m_2,m_3)\ =
\ {1\over 12}(2J_1+1)(N-2J_1) (N+1)(N+1-2J_1)(N+2)\ . \eqno(21)
$$ These two dimensions can coincide only in the exceptional
cases $N=2$, $J_1 = 0$, $d(2,0) = d_c = 6$ and $N = 2J_1+1$ where
we have $$ d_0\equiv d(2J_1+1,J_1) = d_c = {1\over 3}(J_1+1)
(2J_1+1)(2J_1+3) = {1\over 6}N(N+1)(N+2)\ .
\eqno(22)$$

Until now we have discussed the case $j_2 = 0$, $j_1\ge 0$. The
other case, $j_1 = 0$, $j_2\ge 0$, is obtained trivially from
this. In fact, if we introduce the helicity $h = j_1-j_2$, then
all the formulae above may be written in terms of $|h| $; in
particular, for the exceptional case $N = 2 |h| + 1$ we have (cf.
(22)): $$ d_0\ = \ {1\over 3}(| h | +1) (2| h |+1)(2| h |+3)\ = \
{1\over 6} N(N+1)(N+2)\ . \eqno(23) $$

We give now some explicit examples to illustrate the above
classification. Take the massless case $d = j_1+1$, $j_2 = 0$,
and consider the three examples $j_1 = 1/2$ and $j_1 = 1$ for $N
= 3$, and $j_1 = 1/2$ for $N = 2$.

\bigskip

\noindent \line{$\bullet$ $N = 3$, $j_1 = 1/2$, $q = e^{2\pi
i/3}$\hfill} \smallskip

According to (19) and (20) we have: $m_1 = 2$, $m_2 = 2$, $m_3 =
1$, and $d(3,1/2) = 16$. Note that the classical dimension for
this case is $d_c = 20$. The singular vectors corresponding to
the simple roots are : $(X_1^+)^2v_0$, $(X_2^+)^2 v_0$ and
$X_3^+\, v_0$; thus, in the irreducible representation with
vacuum state $| \ \rangle$, we have $$ (X_1^+)^2| \ \rangle =
0,\quad (X_2^+)^2| \ \rangle = 0, \quad X_3^+ | \ \rangle = 0\
.\eqno(24) $$ The sixteen basis states are the vacuum $| \
\rangle$ and $$\eqalign{&X_1^+| \ \rangle\cr &X_3^+ X_2^+| \
\rangle\cr &X_3^+ X_2^+ X_1^+| \ \rangle\cr &X_3^+ X_2^+ X_1^+
X_2^+| \ \rangle\cr &X_2^+ X_3^+ X_2^+ X_1^+ X_2^+ X_1^+| \
\rangle\cr}\quad \eqalign{ &X_2^+| \ \rangle \cr &X_1^+ X_2^+
X_1^+| \ \rangle\cr &X_2^+ X_1^+ X_2^+ X_1^+| \ \rangle\cr &X_3^+
X_2^+ X_1^+ X_2^+ X_1^+| \ \rangle\cr &(X_3^+)^2 X_2^+ X_1^+
X_2^+ X_1^+| \ \rangle \cr}\quad \eqalign{ &X_1^+ X_2^+| \
\rangle \cr &X_2^+ X_1^+ X_2^+| \ \rangle\cr &X_3^+ X_1^+ X_2^+
X_1^+| \ \rangle\cr &(X_3^+)^2 X_2^+ X_1^+ X_2^+| \ \rangle\cr
&X_2^+ (X_3^+)^2 X_2^+ X_1^+ X_2^+ X_1^+| \ \rangle\ .\cr}
\eqno(25)$$ One can explicitly check that the norms with respect
to the ${\cal U}_q(su(2,2))$-invariant scalar product are
positive (in fact, are all equal to 1). All other vectors have
zero-norm and are decoupled from the representation. The same
happens in the other two examples below.

\bigskip

\noindent \line{$\bullet$ $N = 3$, $j_1 = 1$, $q = e^{2\pi i/3}$
\hfill} \smallskip

In this case one has: $m_1 = 3$, $m_2 = 1$, $m_3 = 1$, and $d_0 =
d_c = 10$. Note that now the dimension of the representation
coincides with the classical one. The singular vectors
corresponding to the simple roots are easily obtained: $(X_1^+)^3
v_0$, $X_2^+ v_0$ and $X_3^+ v_0$. In the irreducible
representation one thus has $$ (X_1^+)^3| \ \rangle = 0,\quad
X_2^+| \ \rangle = 0, \quad X_3^+| \ \rangle = 0\ .\eqno(26) $$
The ten states spanning this representation are given by the
vacuum $| \ \rangle$ and $$\eqalign{&X_1^+| \ \rangle\cr &X_1^+
X_2^+ X_1^+| \ \rangle\cr &X_3^+ X_1^+ X_2^+ X_1^+| \
\rangle\cr}\qquad \eqalign{ &(X_1^+)^2| \ \rangle \cr &X_3^+
X_2^+ X_1^+| \ \rangle\cr &X_3^+ (X_2^+)^2 (X_1^+)^2| \
\rangle\cr}\qquad \eqalign{ &X_2^+ X_1^+| \ \rangle \cr &X_2^+
X_1^+ X_2^+ X_1^+| \ \rangle\cr &(X_3^+)^2 (X_2^+)^2 (X_1^+)^2| \
\rangle ~. \cr}\eqno(27)$$ Also in this case one can explicitly
check that the norms with respect to the $\Us$ invariant scalar
product are all equal to unity.

\bigskip

\noindent \line{$\bullet$ $N = 2$, $j_1 = 1/2$, $q = e^{i\pi } =
-1$ \hfill} \smallskip

This is a $q$-deformation of the fundamental representation.
According to (19) and (22) we have now: $m_1 = 2$, $m_2 = 1$,
$m_3 = 1$, and $d_0 = d_c = 4$, so that also in this case the
dimension of the representation is equal to the classical one.
The singular vectors corresponding to the simple roots are
$(X_1^+)^2 v_0$, $X_2^+ v_0$ and $X_3^+ v_0$, and in the
irreducible representation one has $$ (X_1^+)^2| \ \rangle =
0,\quad X_2^+| \ \rangle = 0, \quad X_3^+| \ \rangle = 0\
.\eqno(28) $$ The remaining four basis vectors are given by: $$|
\ \rangle\ ,\quad X_1^+| \ \rangle\ ,\quad X_2^+ X_1^+| \
\rangle\ , \quad X_3^+ X_2^+ X_1^+| \ \rangle\ ,\eqno(29)$$ and
all have unit norm. In this case one can easily work out the
$4\times 4$ matrices representing the generators of ${\cal
U}_q(sl(4,{\bf C}))$. With $$\sigma_+ = \left( \matrix{0&1\cr
0&0} \right) \ ,~~ \sigma_- = \left( \matrix{0&0\cr 1&0} \right)
\ ,~~ \sigma_3 = \left( \matrix{1&0\cr 0&-1} \right) \ ,~~ e_1 =
\left( \matrix{1&0\cr 0&0} \right) \ ,~~ e_2 = \left(
\matrix{0&0\cr 0&1} \right) \ \eqno(30)$$ one finds:
$$\eqalign{&H_1 = - \left( \matrix{\sigma_3 & 0\cr 0 & 0\cr}
\right) \cr &X_1^+ = \left( \matrix{\sigma_-& 0\cr 0 & 0\cr}
\right) \cr &X_1^- = \left( \matrix{\sigma_+& 0\cr 0 & 0\cr}
\right) \cr &X_{12}^+ = -i \left( \matrix{0 & 0\cr e_1 & 0\cr}
\right) \cr &X_{12}^- = -i \left( \matrix{0 & e_1\cr 0 & 0\cr}
\right) \cr}\qquad \eqalign{&H_2 = \left( \matrix{e_2 & 0\cr 0 &
-e_1 \cr} \right) \cr &X_2^+ = \left( \matrix{0 & 0\cr \sigma_+ &
0\cr} \right) \cr &X_2^- = - \left( \matrix{0 & \sigma_-\cr 0 &
0\cr} \right) \cr &X_{23}^+ = i \left( \matrix{0 & 0\cr e_2 & 0}
\right) \cr &X_{23}^- = i \left( \matrix{0 & e_2\cr 0 & 0}
\right) \cr}\qquad \eqalign{&H_3 = - \left( \matrix{ 0& 0\cr 0 &
\sigma_3\cr} \right) \cr &X_3^+ = \left( \matrix{0& 0\cr 0 &
\sigma_-\cr} \right) \cr &X_3^- = \left( \matrix{0& 0\cr 0 &
\sigma_+\cr} \right) \cr &X_{13}^+ = - \left( \matrix{0 & 0\cr
\sigma_- & 0\cr} \right) \cr &X_{13}^- = \left( \matrix{0 &
\sigma_+\cr 0 & 0\cr} \right) \cr} \eqno(31)$$ Note that in this
basis the anti-linear anti-involution $\om$ (cf. (6)) is
represented by matrix hermitean conjugation. The matrix form of
the skew - hermitean generators of $\Us$ can be easily obtained
from the above expressions with the help of Eq. $(7)$. One can
now explicitly check that this representation of $\Us$ is
unitary.

\smallskip

As a final remark, we note that it may be possible to apply the
finite dimensional UIRs given above to the theory and
classification of elementary particles. This is suggestive
because there is a simple relation between the four $\times$ four
representation given above and the fundamental representation
matrices of $su(4)$ (e.g. [10]).

\vskip 1cm

{\bf Acknowledgments}

V.K.D. and V.H. would like to thank Professor Abdus Salam for
hospitality and financial support at the ICTP. V.K.D. was also
partially supported by the Bulgarian National Foundation for
Science, Grant $\Phi - 11$.

\bigskip

{\bf References}

\bigskip

\item{1.} G. Mack, Comm. Math. Phys. {\bf 55} (1977) 1.
\smallskip \item{2.} V.K. Dobrev, Canonical $q$-deformations of
noncompact Lie (super-) algebras, G\"ottingen University
preprint, (July 1991), to appear in J. Phys. A; ~cf. also:
$q$-Deformations of noncompact  Lie (super-) algebras: the
examples of $q$-deformed Lorentz, Weyl, Poincar\'e and (super-)
conformal algebras, ICTP preprint IC/92/13 (1992), to appear in
the Proceedings of the Quantum Groups Workshop of the Wigner
Symposium (Goslar, July 1991). \smallskip \item{3.} V.G.
Drinfeld, Soviet. Math. Dokl. {\bf 32} (1985) 2548; in {\it
Proceedings of the International Congress of Mathematicians},
Berkeley (1986), Vol. 1 (The American Mathematical Society,
Providence, 1987) p.798. \smallskip \item{4.} M. Jimbo, Lett.
Math. Phys. {\bf 10} (1985) 63; Lett. Math. Phys. {\bf 11} (1986)
247. \smallskip \item{5.} E. Abe, {\it Hopf Algebras}, Cambridge
Tracts in Math., N 74, (Cambridge Univ. Press, 1980). \smallskip
\item{6.} V.K. Dobrev, ICTP Trieste internal report IC/89/142
(June 1989) and in {\it Proceedings of the International Group
Theory Conference} (St. Andrews, 1989), Eds. C.M. Campbell and
E.F. Robertson, Vol. 1, London Math. Soc. Lect. Note Ser. 159
(1991) p. 87. \smallskip \item{7.} V.K. Dobrev, Rep. Math. Phys.
{\bf 25} (1988) 159. \smallskip \item{8.} C. De Concini and V.G.
Kac, Progress in Math. {\bf 92} (Birkh\"auser, Boston, 1990) p.
471. \smallskip \item{9.} H.H. Andersen, P. Polo and K. Wen,
Representations of quantum algebras, Aarhus University, Math.
Inst. preprint series 1989/1990 No. 24 (1990). \smallskip
\item{10.} D.B. Lichtenberg, {\it Unitary Symmetry and Elementary
Particles} (2nd ed.), (Academic Press, NY, 1978).
\vfill\eject\end